\documentclass[a4paper,11pt]{article} 
\usepackage[utf8]{inputenc}
\usepackage{authblk}
\usepackage{fullpage}
\usepackage{graphicx}
\usepackage{xcolor}
\usepackage{algorithm}
\usepackage{algpseudocode}
\usepackage{amssymb}
\usepackage{amsthm}
\usepackage{amsmath}
\usepackage{url}
\usepackage{bbold}
\usepackage[square,numbers,comma,sort&compress]{natbib} 
\usepackage[colorlinks,urlcolor=red,citecolor=red,linkcolor=blue,pdftex, pdfauthor = Lukas Exl]{hyperref}
\usepackage{tabularx}
\usepackage{setspace} 
\usepackage{listings}
\lstdefinestyle{customc}{
  belowcaptionskip=1\baselineskip,
  breaklines=true,
  frame=L,
  xleftmargin=\parindent,
  language=MATLAB,
  showstringspaces=false,
  basicstyle=\footnotesize\ttfamily,
  keywordstyle=\bfseries\color{green!40!black},
  commentstyle=\itshape\color{purple!40!black},
  identifierstyle=\color{blue},
  stringstyle=\color{orange},
}

\lstdefinestyle{customasm}{
  belowcaptionskip=1\baselineskip,
  frame=L,
  xleftmargin=\parindent,
  language=[x86masm]Assembler,
  basicstyle=\footnotesize\ttfamily,
  commentstyle=\itshape\color{purple!40!black},
}

\newcommand{\be}{\begin{equation}}
\newcommand{\ee}{\end{equation}}
\newcommand{\ba}{\begin{array}}
\newcommand{\ea}{\end{array}}
\newcommand{\bea}{\begin{eqnarray}}
\newcommand{\eea}{\end{eqnarray}}
\newcommand{\beas}{\begin{eqnarray*}}
\newcommand{\eeas}{\end{eqnarray*}}
\newcommand{\bx}{{\bf x}}
\newcommand{\by}{{\bf y}}

\newcommand{\bc}{{\bf c}}
\newcommand{\bd}{{\bf d}}

\newcommand{\bk}{{\bf k}}
\newcommand{\bB}{{\bf B}}

\title{A GPU accelerated and error-controlled solver for the unbounded Poisson equation in three dimensions}

\author[1,2]{Lukas Exl \thanks{\texttt{lukas.exl@univie.ac.at}}}
\affil[1]{Fak. Mathematik, Univ. Wien, Oskar-Morgenstern-Platz 1, 1090, Vienna.} 
\affil[2]{Inst. of Solid State Physics, TU Wien, Karlsplatz 13, 1040, Vienna.} 
\begin{document}
\maketitle
\textbf{Abstract.} 
An efficient solver for the three dimensional free-space Poisson equation is presented.  The underlying numerical method is based on finite Fourier series approximation. 
While the error of all involved approximations can be fully controlled, the overall computation error is driven by the convergence of the finite Fourier series of the density. 
For smooth and fast-decaying densities the proposed method will be spectral accurate. The method scales with $\mathcal{O}(N\log N)$ operations, where $N$ is the total 
number of discretization points in the Cartesian grid. The majority of the computational costs come from fast Fourier transforms (FFT), which makes it ideal for GPU computation. 
Several numerical computations on CPU and GPU validate the method and show efficiency and convergence behavior. Tests are performed using the Vienna Scientific Cluster 3 (VSC3).    
A free MATLAB implementation for CPU and GPU is provided to the interested community. \\

dipole-dipole potential, separable Gaussian-sum (GS) approximation

\section{Introduction}
   
The purpose of this paper is to provide the interested reader with a MATLAB implementation of an efficient and mathematically analyzed method \cite{exl2016accurate} for solving the free-space/unbounded Poisson equation. 
More precisely, the method presented in this paper solves 
\begin{align}\label{eqn:problem1}
 - \Delta u(\bx) = \rho(\bx), \quad \bx \in \mathbb{R}^3, \quad \lim_{|\bx| \rightarrow \infty} |u(\bx)| = 0,
\end{align}
via the well-known representation of the solution to \eqref{eqn:problem1} as the convolution of the density $\rho$ with the free-space Green's function $U(\bx) = \tfrac{1}{4\pi} \tfrac{1}{|\bx|}$ 
\begin{align}\label{eqn:conv}
 u(\bx) = (U \ast \rho)(\bx) = \int_{\mathbb{R}^3} U(\bx - \by)\rho(\by)d\by,\quad \bx \in \mathbb{R}^3.
\end{align}
The problem \eqref{eqn:problem1} is fundamental in many fields of physics, e.g. 
quantum chemistry \cite{leach2001molecular,martyna1999rec,Wavelet06,Wavelet07,fusti2002accurate,bao2010efficient},   
particle physics \cite{arnold2005efficient,hejlesen2016multiresolution} or astrophysics \cite{budiardja2011parallel}.
Therefore, the provided implementation might serve as improvement of existing simulation codes that use the high-level computing environment MATLAB. 
However, the provided code could also be understood as an easily readable open source prototype, ready for translation to other different programming languages. \\

In the following Sec.~\ref{method} the method is described mathematically, followed by a section about computational aspects and approximation errors (see Sec.~\ref{compaspects}). Sec.~\ref{usage} describes the usage of 
the implementation by means of a test example.  
Validation of the implementation and tests for computational efficiency (see Sec.~\ref{numerics}) show practical applicability. Test runs are performed on the Vienna Scientific Cluster 3 (VSC3) both on 
CPU nodes and Tesla GPU devices.

\section{Method description}\label{method}
The method of this paper is in the class of Ewald type methods \cite{ewald1921berechnung,heyes1981electrostatic,martyna1999rec}.
Those approaches split the singular convolution kernel $U$ into a smooth long-range part $U_s$ and a singular short-range correction $U_c$. The smooth part of the convolution can then be treated with 
help of the convolution theorem, i.e., $U \ast \rho = \mathcal{F}^{-1}(\mathcal{F}(U_{s})\cdot \mathcal{F}(\rho))$. This is usually done on an equispaced Cartesian grid with the help of the quasi linearly scaling fast Fourier transform. Here, the smoothness usually leads to fast converging Fourier series, which make the discrete approximation accurate even on 
coarser grids. However, the correction $U_c$ still contains a singularity but is also localized, hence, can be treated with a direct summation approach. Wile a direct evaluation of the convolution \eqref{eqn:problem1} would scale 
with $\mathcal{O}(N^2)$ on a Cartesian grid with a total number of $N$ grid points, the original Ewald method \cite{ewald1921berechnung} and parameter tuned variations of it scale with $\mathcal{O}(N^{3/2})$ operations.
The method described here scales with $\mathcal{O}(N \log N)$ operations. This is achieved by FFT for both parts, the smooth convolution and the correction. The smooth kernel consists of a product of one-dimensional exponential functions 
(Gaussian-sum), which allows the usage of highly accurate one-dimensional adaptive quadrature for computation of the interaction kernel in Fourier space. Taylor expansion of the density in the near-zone allows to treat the correction by 
analytical integration, where involved derivatives are computed by FFT as well. The method is efficient and mathematically proven to yield full control over the maximum computation error \cite{exl2016accurate}.\\
We now give a brief description of the method, where we also emphasize novel aspects relevant to the implementation. 
A detailed mathematical description of the method including error analysis was recently published by the author \cite{exl2016accurate}. 
We adapt it here for the case of general rectangular computational domains.\\ 
The computational box coincides with the domain of target points $\bx$, where the potential $u$ is computed. 
The smooth density $\rho$ is assumed to vanish (up to double precision) outside the computational box, so it is expected to be fast decaying. Our method makes use of a finite Fourier series approximation of the density, which is assumed to be fast converging due to 
the smoothness and compact support of $\rho$. The analysis in \cite{exl2016accurate} assumes the computational domain to be the unit square box $\bB_1 := [-1,1]^3$. 
To generalize the method's framework for the user's convenience, assume the density $\rho$ to be compactly supported in the general rectangular box
$\bB := [a_1,b_1] \times [a_2,b_2] \times [a_3,b_3]$. We will first derive a 'standardized form' of the problem, which allows us to treat the key-approximations of the method 
independently of the concrete choice of the computational box.  
Let $\bc$ be the center of $\bB$ such that $\widetilde{\bB} := \bB - \bc$ is a rectangular box centered at the origin. 
As a consequence of the compact support of the density, the convolution integral \eqref{eqn:conv} is actually over the domain $\bB$. Hence, we can write  
\begin{align}
 u(\bx) = \int_{\widetilde{\bB}} U(\widetilde{\bx} - \widetilde{\by})\rho(\widetilde{\by}+\bc)d\widetilde{\by}, \quad \bx = \widetilde{\bx} + \bc \in \bB.
\end{align}
Now we define $\lambda := \max_{q=1,2,3}\{\tfrac{b_q-a_q}{2} \}$ and $\bB_{1,\lambda} := \tfrac{1}{\lambda}\widetilde{\bB} \subseteq \bB_1$. We get 
\begin{align}
 u(\bx) = \lambda^{2}\int_{\bB_{1,\lambda}} U(\bx^\prime - \by^\prime)\rho(\lambda \by^\prime +\bc)d\by^\prime, \quad \bx = \lambda\bx^\prime + \bc \in \bB.
\end{align}
Changing variables and extending the domain of integration to $2\cdot\bB_{1,\lambda} \supseteq \bB_{1,\lambda} - \bx^\prime,\,\,\bx^\prime \in \bB_{1,\lambda}$ leads to 
\begin{align}\label{eqn:stdform}
 u(\bx) = \lambda^{2}\int_{2\cdot\bB_{1,\lambda}} U(\by)\rho_{\lambda;\bc}(\bx^\prime - \by)d\by, \quad \bx = \lambda\bx^\prime + \bc \in \bB,
\end{align}
where $\rho_{\lambda;\bc}(\bx) := \rho(\lambda\bx + \bc)$ with support in $\bB_{1,\lambda}$.     
The key idea of the method is to approximate the singular kernel $U(\bx) = \tfrac{1}{4\pi} \tfrac{1}{|\bx|}$ with a Gaussian-sum in a region contained in the 
integration domain $2\cdot\bB_{1,\lambda}$ but excluding a $\delta$-ball around the origin (where 
$U$ is singular). The latter step is compensated by a near zone correction. More precisely, for $\delta \in (0,\min_{q=1,2,3}\{\tfrac{b_q-a_q}{2\lambda}\})$ and $\bx = \lambda\bx^\prime + \bc \in \bB$ we get 
\begin{align}\label{eqn:approx}
 u(\bx) \approx \lambda^{2} \,\Big(\int_{2\cdot\bB_{1,\lambda}} U_{GS}(\by)\rho_{\lambda;\bc}(\bx^\prime - \by)d\by + \int_{\bB_{\delta}} (U-U_{GS})(\by)\rho_{\lambda;\bc}(\bx^\prime - \by)d\by\Big) =: \lambda^2\big(I_1(\bx) + I_\delta(\bx)\big),
\end{align}
where $U_{GS}(\by) = \sum_{j=0}^S w_j e^{-\tau_j^2 |\by|^2} =  \sum_{j=0}^S w_j \prod_{q=1}^3 e^{-\tau_j^2 y_q^2} \approx U(\by), \,\,|\by| \in [\delta,2]$ is a Gaussian-sum (GS) approximation 
realized by sinc-quadrature \cite{HackBush2005,exl_2014,exl2016accurate}. The integrand in $I_1(\bx)$ is smooth and its convolution kernel is separable (product of 1d functions). 
Hence, it can be treated efficiently by an Fourier based approach. More precisely, it is computed by the inverse Fourier transform of the product of the Fourier transform of the density with the $G$-tensor 
\begin{align}
G_\bk = \sum_{j=0}^S w_j G_\bk^j \quad \text{with}\,\, G_\bk^j = G_{k_1 k_2 k_3}^l = \prod_{q=1}^3 \int_0^2 2 l_q\,e^{-(\tau_j l_q)^2 y_q^2} \cos(\frac{\pi }{2} k_q y_q) \,dy_q, 
\end{align}
where $\bB_{1,\lambda} = [-l_1,l_1] \times [-l_2,l_2] \times [-l_3,l_3]$. The $G$-tensor can be computed accurately by one-dimensional adaptive Gauss-Kronrod quadrature in a setup phase. 
The two Fourier transforms in the (run-time) computation of $I_1$ are efficiently implemented via the FFT with zero-padding, which increases the effort by a factor of eight.\\ 
The correction integral $I_\delta$ is calculated by inserting the third order Taylor polynomial of the shifted density 
$\rho_{\lambda;\bc;\bx^\prime}(\by):=\rho_{\lambda;\bc}(\bx^\prime-\by)$ around $\boldsymbol{0}$, followed by analytical integration in spherical coordinates. The contributions of odd derivatives in the Taylor expansion and 
the off-diagonal elements of the Hessian cancel out. The remaining derivatives are computed from the finite Fourier series of the density, which makes it a scalar multiplication. 
This step is realized by using forward and backward FFT.
\section{Computational aspects and approximation errors}\label{compaspects}
For the concrete computation the computational box is discretized equidistantly with $N := n_1 n_2 n_3$ Cartesian grid points, i.e., the $q$-th principal direction is discretized equidistantly with $n_q$ points. 
The solver's setup phase consists mainly of the precomputation of the $G$-tensor.  
The two Fourier transforms (one forward, one backward) in the computation of $I_1$ in \eqref{eqn:approx} are of size $8N$ and scale with $\mathcal{O}(8N\,\log 8N)$ operations utilizing the FFT. 
The evaluation of the near zone correction makes use of two FFTs (one forward, one backward) of size $N$ and therefore scales with $\mathcal{O}(N\,\log N)$ operations. 
Other operations (multiplications and additions) contribute with linear scaling $\mathcal{O}(N)$.\\ 
Besides the error coming from the finite Fourier series approximation of the density, the (maximum-) error of the convolution method   
is (i) in the computation of $I_1$ due to the Gaussian-sum approximation in $[\delta,2]$ and (ii) in the computation of the correction $I_\delta$ due to the Taylor expansion of the density. 
For fixed $\delta$ the error (i) is controlled by a parameter $\epsilon>0$ in the order of around machine precision, 
while the error (ii) amounts to $\delta^6$. The overall maximum-error of the involved approximations is therefore in the order of $\max\{\epsilon,\delta^6\}$. In practice, $\delta$ will be around $0.005-0.001$, hence, 
yielding an overall maximum-error of the involved approximations of around $\epsilon$. Thus, the overall computation error can be expected to be determined by the convergence of the finite Fourier series of the density (spectral accuracy). 
It is also known from the error analysis in \cite{exl2016accurate} that the approximations of the method without the correction $I_\delta$ yield a maximum-error in the order of $\max\{\delta^2,\epsilon\} \sim 10^{-6}$. 
If the error coming from the finite Fourier series lies above this threshold, the correct $I_\delta$ will lead to no improvement. For coarse grids the error from the Fourier series can be expected to exceed the threshold, such that 
$\delta$ does not have to be chosen too small. This reflects in the computation time of the setup phase, since a larger choice of $\delta$ leads to a smaller number $S$ of terms in the precomputation of 
the $G$-tensor. The heuristic choice of $\delta:= \min \tfrac{b_q-a_q}{50 n_q} \in (0,\min_{q=1,2,3}\{\tfrac{b_q-a_q}{2\lambda}\})$ for $\lambda \leq 25 \min_q n_q$ depends on the discretization size according to the just mentioned considerations. 
However, we take $\delta \geq 10^{-3}$ as a minimum threshold, since the $I_\delta$-correction yields accuracy in the order of $\delta^6$.   
\section{Usage of the solver}\label{usage}
Usage of the solver is simple, see Listing~\ref{lst:usage}. 
The user defines the computational box $\bB$ and its discretization by uniformly discretized edges. 
Next the GPU flag is set and the setup of the solver is accomplished. In Listing~\ref{lst:usage} a Gaussian test density with compact support in $\bB$ is chosen which is sampled on 
the discretized computational box. The actual computation is performed by the solver's \texttt{solve} method. Afterwards the solver could be reused without renewed setup, e.g.  
in large simulations where the potential has to be computed several times on the same geometry. In Listing~\ref{lst:usage} the error computation is demonstrated as well, which is only possible for the analytically given test density.  

\lstset{style=customc,caption={Usage of the \texttt{GSPoisson3d} solver.},label = {lst:usage}}
\begin{lstlisting}
%==============================================================
% Setup geometry
%==============================================================

% Define the computational box
x_min = -2.0;
y_min = -2.0;
z_min = -2.0;

x_max = +2.0;
y_max = +2.0;
z_max = +2.0;

% Number of discretization points for each principal direction 
Jx = 2^6;
Jy = 2^6;
Jz = 2^6;

% Uniformly discretized axes
x = linspace(x_min, x_max, Jx);
y = linspace(y_min, y_max, Jy);
z = linspace(z_min, z_max, Jz);

%==============================================================
% Setup solver
%==============================================================

% GPU flag
use_gpu = false; % true;

% setup solver
solver = GSPoisson3d(x, y, z, use_gpu);

%==============================================================
% Density
%==============================================================

% example density 
[f, u_ref] = problems.gaussian(x, y, z, 0.2);

if use_gpu == true
   f     = gpuArray(f);
   u_ref = gpuArray(u_ref);
end

%==============================================================
% Actual computation
%==============================================================

tic;
u = solver.solve(f);
time = toc;

% Error
E = max( abs( u(:) - u_ref(:) ) ) / max( abs( u_ref(:) ) );

if use_gpu == true
   E = gather(E);
end

fprintf('Maximum error: %.4e\n', E)
fprintf('Time: %.3f (s)\n',time)
\end{lstlisting}

\section{Numerical validation}\label{numerics}
We test our solver for different choices of the density and give maximum relative errors $E$ according to 
\begin{align}\label{error_infty}
E:=\frac{\|u-u_{\vec h}\|_{l^\infty}}{\|u\|_{l^\infty}}=
\frac{\max_{\bx\in \mathcal{T}_h} |u(\bx)-u_{\vec h}(\bx)|}{\max_{\bx\in \mathcal{T}_h} |u(\bx)|},
\end{align}
where $\mathcal{T}_h$ is the rectangular computational domain discretized uniformly in each direction with mesh sizes $\vec h = (h_x,h_y,h_z)^T$. Errors and computation times are compared for the CPU and the GPU case. 
In the following we denote our solver with \texttt{GSPoisson3d} solver. 
The computations were submitted jobs on the Vienna Scientific Cluster 3 (VSC3) which consists of nodes with Intel Xeon E5-2650v2 2.6GHz processors and Tesla K20m GPU devices. 
To accurately give the timings we measure the average times of $100$ computations.

\subsection{Gaussian source}
First we test with the Gaussian density 
\begin{align}\label{eqn:gaussian}
 \rho(\bx) = \frac{1}{(2\pi)^{3/2}\sigma^3 } e^{-|\bx-\bc|^2/(2\sigma^2)}, 
\end{align}
where $\bc \in \mathbb{R}^3$ is the center of the computational box. The exact solution is known to be
\begin{align}
 u^\ast(\bx) = \frac{1}{4 \pi |\bx-\bc|} \textrm{Erf}\Big(\frac{|\bx-\bc|}{\sqrt{2}\sigma}\Big). 
\end{align}
We vary the shape parameter $\sigma$ in our tests and compare errors and computation times on CPU and GPU, 
see Tab.~\ref{tab:gaussian} for the computational domain $\bB=[-2,2]^3$.  

\begin{table}[h!]
\tabcolsep 0pt \caption{Errors and timings for Gaussian density \eqref{eqn:gaussian} in $[-2,2]^3$. Errors $E$, times $t_{\texttt{cpu}}$ and $t_{\texttt{gpu}}$ on CPU and GPU respectively.} \label{tab:gaussian}
\begin{center}\vspace{-1em}
\def\temptablewidth{14cm}
{\rule{\temptablewidth}{1.2pt}}
\begin{tabularx}{14cm}{X X X X X}
$\sigma$ & $N$     & $E$  & $t_{\texttt{cpu}}$ & $t_{\texttt{gpu}}$ \\ \hline
$0.20$   & $16^3$  & 1.659E-03	        & 6.80E-03                       & 1.20E-02                       \\
$0.20$   & $32^3$  & 4.154E-09	        & 1.91E-02                       & 6.27E-03                       \\
$0.20$   & $64^3$  & 6.197E-16	        & 1.17E-01                       & 1.17E-02                       \\
$0.20$   & $128^3$ & 1.052E-15	        & 7.95E-01                       & 5.29E-02                      \\\hline

$0.15$   & $16^3$  & 2.986E-02	        & 4.37E-03                       & 5.44E-03                      \\
$0.15$   & $32^3$  & 2.937E-06	        & 1.80E-02                       & 6.26E-03                      \\
$0.15$   & $64^3$  & 9.386E-16	        & 1.17E-01                       & 1.20E-02                      \\
$0.15$   & $128^3$ & 1.187E-15	        & 8.66E-01                       & 5.26E-02                      \\\hline

$0.10$   & $16^3$  & 3.802E-01	        & 4.34E-03                       & 6.39E-03                       \\
$0.10$   & $32^3$  & 1.129E-03	        & 1.91E-02                       & 6.26E-03                      \\
$0.10$   & $64^3$  & 2.624E-09	        & 1.18E-01                       & 1.10E-02                       \\
$0.10$   & $128^3$ & 1.593E-15	        & 8.22E-01                       & 5.26E-02                       \\\hline
\end{tabularx}
{\rule{\temptablewidth}{1pt}}
\end{center}
\end{table}

\begin{figure}[hbtp]
\hspace{0.0 in} 
\centering 
\includegraphics[scale=0.35]{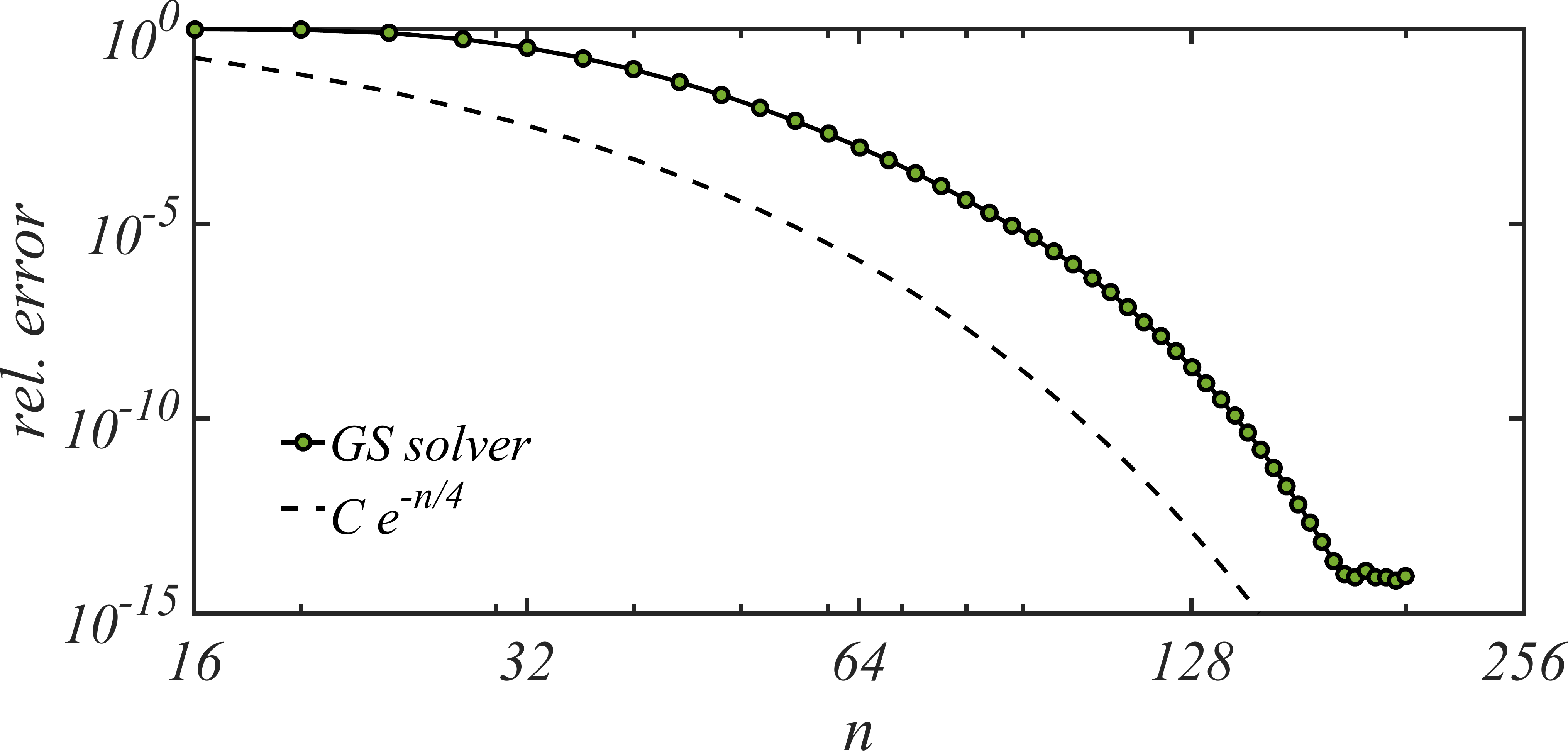}
\caption{Errors for the density in \eqref{eqn:gaussian} in $\bB = [-2,2]^3$ with $\sigma = 0.05$.}\label{fig:conv}
\end{figure}

Fig.~\ref{fig:conv} shows the convergence of the method for example~\eqref{eqn:gaussian} with $\sigma = 0.05$ in $\bB = [-2,2]^3$ .  

\begin{figure}[hbtp]
\hspace{0.0 in} 
\centering 
\includegraphics[scale=0.35]{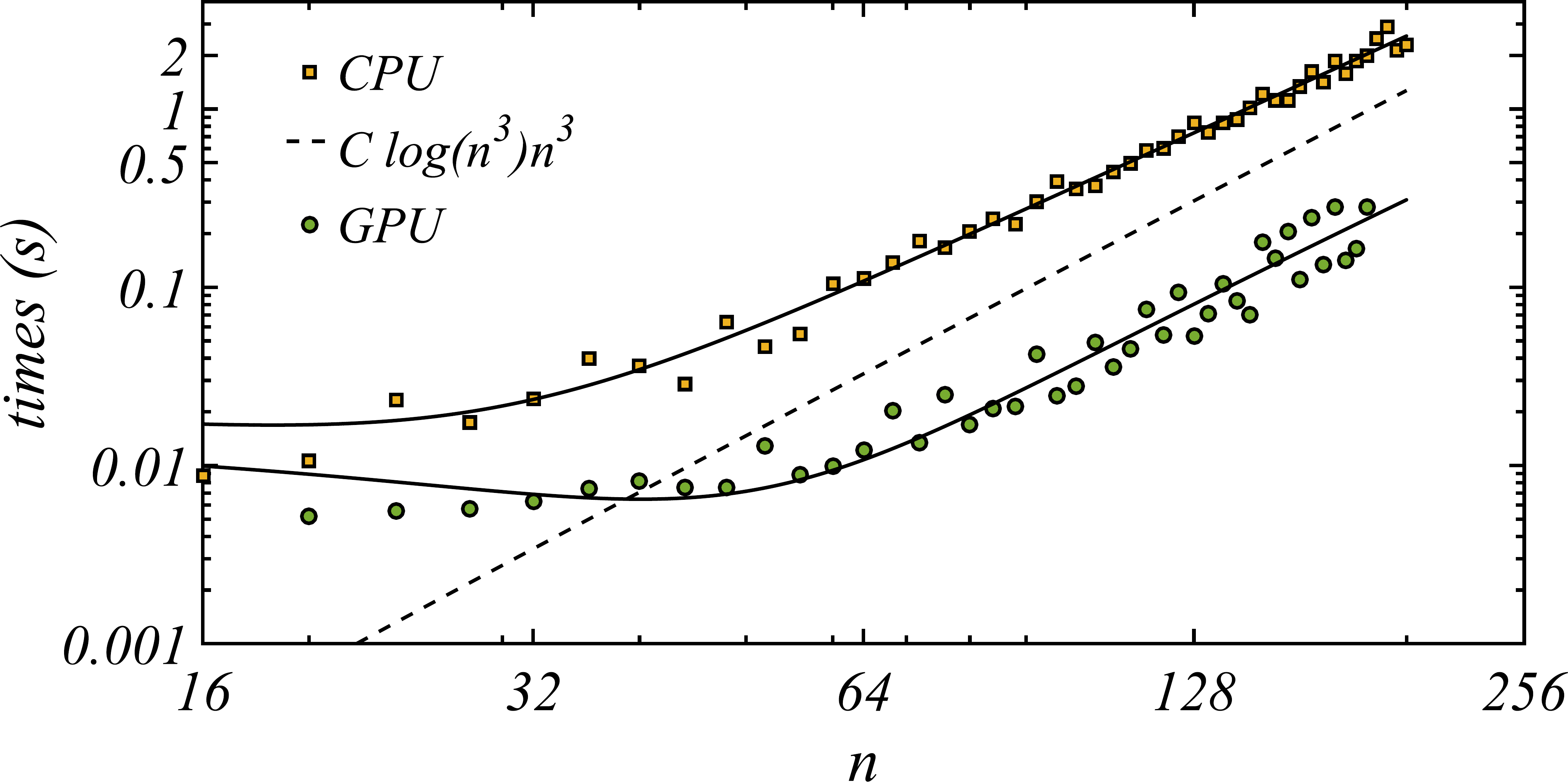}
\caption{Computation times for CPU and GPU on the Vienna Scientific Cluster 3 (VSC3) for the density in \eqref{eqn:gaussian} in $\bB = [-2,2]^3$ with $\sigma = 0.05$.}\label{fig:times}
\end{figure}

This examples show that the \texttt{GSPoisson3d} solver converges up to exponentially fast.\\
Fig.~\ref{fig:times} compares computation times for CPU and GPU for the computations associated with those of Fig.~\ref{fig:conv}.  
Computation times show the $N\log N$ scaling. The GPU acceleration yields a speed up of up to $10$ in the case of larger $N$.\\

Finally, Tab.~\ref{tab:gaussian2} shows results for the rectangular computational domain $\bB=[-3,2] \times [-2,3.5] \times [-1,5]$ and $\sigma=0.2$. 

\begin{table}[h!]
\tabcolsep 0pt \caption{Errors and timings for Gaussian density \eqref{eqn:gaussian} in $[-3,2] \times [-2,3.5] \times [-1,5]$ and $\sigma=0.2$. Errors $E$, times $t_{\texttt{cpu}}$ and $t_{\texttt{gpu}}$ on CPU and GPU respectively.  } \label{tab:gaussian2}
\begin{center}\vspace{-1em}
\def\temptablewidth{14cm}
{\rule{\temptablewidth}{1.2pt}}
\begin{tabularx}{14cm}{ X X X X}
 $N$     & $E$  & $t_{\texttt{cpu}}$ & $t_{\texttt{gpu}}$ \\ \hline
 $16^3$  & 4.417E-02	        & 6.65E-03                       & 1.02e-02                      \\
 $32^3$  & 1.857E-05	        & 1.89E-02                       & 7.58e-03                      \\
 $64^3$  & 3.126E-12	        & 1.29E-01                       & 1.21e-02                       \\
 $128^3$ & 1.643E-12	        & 8.18E-01                       & 5.28e-02                    \\\hline
\end{tabularx}
{\rule{\temptablewidth}{1pt}}
\end{center}
\end{table}

\subsection{Superposition of Gaussian sources}
We test with the Gaussian density 
\begin{align}\label{eqn:gaussians}
 \rho(\bx) = \frac{1}{2}\Big( \frac{1}{(2\pi)^{3/2}\sigma_1^3 } e^{-|\bx-(\bc+\bd)|^2/(2\sigma_1^2)} + \frac{1}{(2\pi)^{3/2}\sigma_2^3 } e^{-|\bx-(\bc-\bd)|^2/(2\sigma_2^2)} \Big), 
\end{align}
where $\bc,\bd\in \mathbb{R}^3$ is the center of the computational box and a shift, respectively. The exact solution is 
\begin{align}
 u^\ast(\bx) =  \frac{1}{2}\Big(\frac{1}{4 \pi |\bx-(\bc+\bd)|} \textrm{Erf}\Big(\frac{|\bx-(\bc+\bd)|}{\sqrt{2}\sigma_1}\Big) + \frac{1}{4 \pi |\bx-(\bc-\bd)|} \textrm{Erf}\Big(\frac{|\bx-(\bc-\bd)|}{\sqrt{2}\sigma_2}\Big)\Big). 
\end{align}
We compare errors and computation times on CPU and GPU and give the results in Tab.~\ref{tab:gaussians}.

\begin{table}[h!]
\tabcolsep 0pt \caption{Errors and timings for Gaussian density \eqref{eqn:gaussians} in $[-2,2]^3$ with $\bd = (0.1,-0.05,0.05)^T$ and $\sigma_1 = 0.2$ and $\sigma_2 = 0.1$. 
Errors $E$, times $t_{\texttt{cpu}}$ and $t_{\texttt{gpu}}$ on CPU and GPU respectively.  } \label{tab:gaussians}
\begin{center}\vspace{-1em}
\def\temptablewidth{14cm}
{\rule{\temptablewidth}{1.2pt}}
\begin{tabularx}{14cm}{X X X X}
$N$     & $E$  & $t_{\texttt{cpu}}$ & $t_{\texttt{gpu}}$ \\ \hline
$16^3$  & 5.663E-02	        & 1.14E-02                       &       2.12e-02              \\
$32^3$  & 1.533E-03	        & 2.26E-02                       &         6.65e-03             \\
$64^3$  & 5.920E-09	        & 1.30E-01                       &        1.17e-02               \\
$128^3$ & 1.246E-15	        & 8.72E-01                       &        5.26e-02               \\\hline
\end{tabularx}
{\rule{\temptablewidth}{1pt}}
\end{center}
\end{table}

\subsection{Bump function}
Next we consider the following bump function as density ($d,R>0$)
\begin{equation}\label{eqn:bump}
 \rho_{R,c}(\bx) = 
 \left\{\begin{array}{ll}
  2d R^2 \,\frac{3R^4-2R^2 |\bx-\bc|^2 - |\bx-\bc|^4 - d R^2 |\bx-\bc|^2 }{(R^2-|\bx-\bc|^2 )^{4}}\,  e^{-d R^2\,(R+|\bx-\bc|) / (R - |\bx-\bc|)}, & |\bx-\bc|<R\\
  0, & |\bx-\bc|\geq R,
 \end{array}\right.
\end{equation}
where the exact solution is given as
\begin{align}
 u^\ast(\bx) = 
 \left\{\begin{array}{ll}
         e^{\tfrac{-d}{1-|\bx-\bc|^2 /R^2}} & |\bx-\bc|<R\\
         0, & |\bx-\bc|\geq R.
        \end{array}\right.
\end{align}

\begin{table}[h!]
\tabcolsep 0pt \caption{Errors and timings for the density in \eqref{eqn:bump} in $[-3,1] \times [-2,3] \times [-2,4]$ with $d = 10$ and $R = 2$. Errors $E$, times $t_{\texttt{cpu}}$ and $t_{\texttt{gpu}}$ on CPU and GPU respectively.  } \label{tab:bump}
\begin{center}\vspace{-1em}
\def\temptablewidth{14cm}
{\rule{\temptablewidth}{1.2pt}}
\begin{tabularx}{14cm}{X X X X}
$N$     & $E$  & $t_{\texttt{cpu}}$ & $t_{\texttt{gpu}}$ \\ \hline
$16^3$  & 2.070E-03	     & 6.64E-03		        & 1.05E-02                \\
$32^3$  & 3.928E-06	     & 2.26E-02         	& 7.76E-03                \\
$64^3$  & 9.264E-10	     & 1.19E-01       		& 1.18E-02                \\
$128^3$ & 4.973E-13	     & 7.95E-01       		& 5.26E-02                \\\hline
\end{tabularx}
{\rule{\temptablewidth}{1pt}}
\end{center}
\end{table}

Tab.~\ref{tab:bump} shows the results in the rectangular domain $[-3,1] \times [-2,3] \times [-2,4]$ with $d = 10$ and $R = 2$.

\subsection{Anisotropic Gaussian}
We test for the anisotropic density ($\bc = (c_x,c_y,c_z)^T,\,\,\sigma_x,\sigma_y,\sigma_z > 0$)
\begin{align}\label{eqn:anis}
  \rho(\bx) =  -( \frac{4(x-c_x)^2}{\sigma_x^4} + \frac{4(y-c_y)^2}{\sigma_y^4} + \frac{4(z-c_z)^2}{\sigma_z^4} - \frac{2}{\sigma_x^2}- \frac{2}{\sigma_y^2}- \frac{2}{\sigma_z^2} )\,e^{-(x-c_x)^2/\sigma_x^2 -(y-c_y)^2/\sigma_y^2 -(z-c_z)^2/\sigma_z^2}, 
\end{align}
which is produced by Eqn.~\eqref{eqn:problem1} and the prescribed exact solution 
\begin{align}
 u^\ast(\bx) = e^{-(x-c_x)^2/\sigma_x^2 -(y-c_y)^2/\sigma_y^2 -(z-c_z)^2/\sigma_z^2}.
\end{align}
Tab.~\ref{tab:anis} shows the results for $\boldsymbol{\sigma} = (0.30,0.20,0.28)^T$ on $\bB = [-2,2]^3$.\\

\begin{table}[h!]
\tabcolsep 0pt \caption{Errors and timings for the density in \eqref{eqn:anis} in $[-2,2]^3$ with $\boldsymbol{\sigma} = (0.30,0.20,0.28)^T$. Errors $E$, times $t_{\texttt{cpu}}$ and $t_{\texttt{gpu}}$ on CPU and GPU respectively. } \label{tab:anis}
\begin{center}\vspace{-1em}
\def\temptablewidth{14cm}
{\rule{\temptablewidth}{1.3pt}}
\begin{tabularx}{14cm}{X X X X}
$N$     & $E$  & $t_{\texttt{cpu}}$ & $t_{\texttt{gpu}}$ \\ \hline
$16^3$  & 4.208E-01	        & 6.51E-03                       & 1.73E-02                      \\
$32^3$  & 1.627E-04	        & 2.07E-02                       & 8.54E-03                       \\
$64^3$  & 1.466E-13	        & 1.24E-01                       & 1.17E-02                       \\
$128^3$ & 1.349E-15	        & 8.01E-01                       & 5.29E-02                      \\\hline
\end{tabularx}
{\rule{\temptablewidth}{1pt}}
\end{center}
\end{table}

\begin{table}[h!]
\tabcolsep 0pt \caption{Errors and timings for the density in \eqref{eqn:anis} in $\bB = [-2,2] \times  [-4,4] \times  [-6,6]$ with $\boldsymbol{\sigma} = (0.10,0.20,0.3)^T$. Errors $E$, times $t_{\texttt{cpu}}$ and $t_{\texttt{gpu}}$ 
on CPU and GPU respectively. } \label{tab:anis2}
\begin{center}\vspace{-1em}
\def\temptablewidth{14cm}
{\rule{\temptablewidth}{1.3pt}}
\begin{tabularx}{14cm}{X X X X}
$N$     & $E$  & $t_{\texttt{cpu}}$ & $t_{\texttt{gpu}}$ \\ \hline
$16^3$  & 1.011e+02        & 6.51E-03                       & 1.73E-02                      \\
$32^3$  &  9.529e-01	   & 2.07E-02                       & 7.54E-03                       \\
$64^3$  & 1.544e-04 	   & 1.24E-01                       & 1.16E-02                       \\
$128^3$ & 7.839e-09	   & 8.01E-01                       & 5.29E-02                      \\
$162^3$ & 8.105e-09        & 1.22E+00 		 	    & 1.08E-01 \\\hline
\end{tabularx}
{\rule{\temptablewidth}{1pt}}
\end{center}
\end{table}   

\begin{figure}[hbtp]
\hspace{0.0 in} 
\centering 
\includegraphics[scale=0.35]{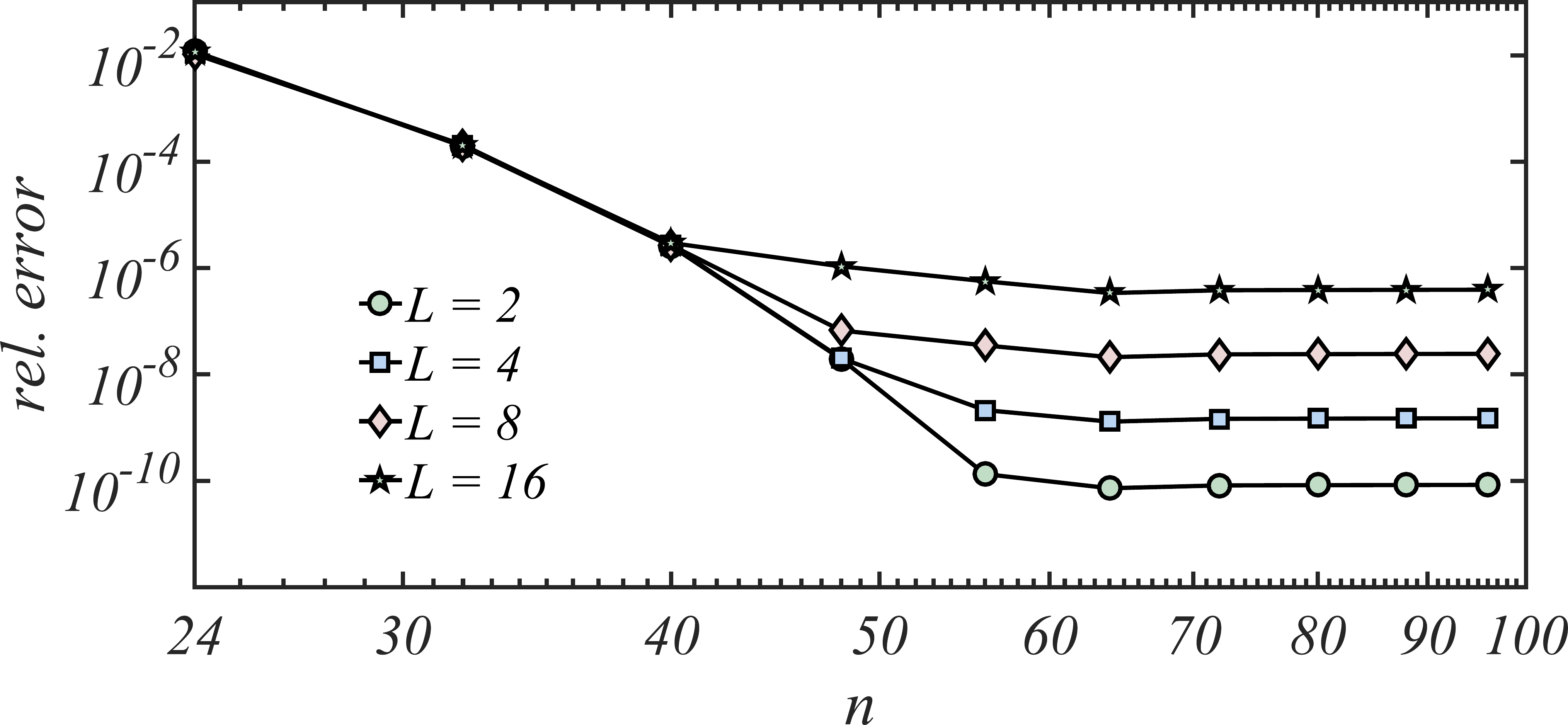}
\caption{Errors for the density in \eqref{eqn:anis} in $\bB = [-2,2] \times [-2L,2L]^2$ for $L = 2,4,8,16$  with
$\boldsymbol{\sigma} = (0.20,0.20\,L,0.20\,L)^T$.}\label{fig:anis3}
\end{figure}

The next experiment takes $\boldsymbol{\sigma} = (0.10,0.20,0.30)^T$ and a rectangular domain $\bB = [-2,2] \times [-4,4] \times  [-6,6]$, which is adjusted to the $\sigma-$values. Tab.~\ref{tab:anis2} shows the results for different mesh sizes. 
Convergence stagnates here at an error level of around $1$e-$9$ most likely due to significant loss of digits in the computation of the $G$-tensor.   
However, large aspect ratios are known to be difficult cases. In fact, the next example shows that for flat domains $\bB = [-2,2] \times [-2L,2L]^2$ for $L = 2,4,8,16$ and adjusted 
$\boldsymbol{\sigma} = (0.20,0.20\,L,0.20\,L)^T$ the error stagnates and convergence gets worse for increasing $L$, compare with Fig.~\ref{fig:anis3}. 
Tests on prolongated domains $\bB = [-2,2]^2 \times [-2L,2L]$ show qualitatively comparable results.
However, already relatively coarse discretizations yield still acceptable error levels if the rectangular domain is not too flat or prolongated.

\subsection{Oscillating density}
We now test for the oscillating density ($\sigma,\omega > 0$)
\begin{align}\label{eqn:osci}
  \rho(\bx) = & \,e^{-|\bx-\bc|^2/\sigma^2} \,( 6\omega \sin(\omega |\bx-\bc|^2) - 4\omega^2 |\bx-\bc|^2 \cos(\omega |\bx-\bc|^2))\\
  & \,- (\frac{4|\bx-\bc|^2}{\sigma^4} - \frac{6}{\sigma^2})\, e^{-|\bx-\bc|^2/\sigma^2}\,\cos(\omega\,|\bx-\bc|^2) 
  -  \frac{8\omega |\bx-\bc|^2}{\sigma^2}\,e^{-|\bx-\bc|^2/\sigma^2}\,\sin(\omega\,|\bx-\bc|^2)
  \end{align}
which is produced by Eqn.~\ref{eqn:problem1} and the prescribed exact solution 
\begin{align}
 u^\ast(\bx) = e^{-|\bx-\bc|^2/\sigma^2}\,\cos(\omega\,|\bx-\bc|^2).
\end{align}
Tab.~\ref{tab:osci} shows the results for $\sigma = 0.30$ and $\omega = 20$ on $\bB = [-2,2]^3$.

\begin{table}[h!]
\tabcolsep 1pt \caption{Errors and timings for the density in \eqref{eqn:osci} in $[-2,2]^3$ with $\sigma = 0.30$ and $\omega = 20$. Errors $E$, times $t_{\texttt{cpu}}$ and $t_{\texttt{gpu}}$ on CPU and GPU respectively.} \label{tab:osci}
\begin{center}\vspace{-1em}
\def\temptablewidth{14cm}
{\rule{\temptablewidth}{1.3pt}}
\begin{tabularx}{14cm}{X X X X}
$N$     & $E$  & $t_{\texttt{cpu}}$ & $t_{\texttt{gpu}}$ \\ \hline
$16^3$  & 6.179E+00	        & 6.76E-03                       & 7.93E-03                      \\
$32^3$  & 7.921E-03	        & 1.85E-02                       & 7.53E-03                      \\
$64^3$  & 3.631E-08	        & 1.30E-01                       & 1.17E-02                      \\
$128^3$ & 2.127E-15             & 8.24E-01                       & 5.26E-02                      \\\hline
\end{tabularx}
{\rule{\temptablewidth}{1pt}}
\end{center}
\end{table}

As expected, the \texttt{GSPoisson3d} solver converges spectrally accurate for the oscillating density.

\section{Conclusions}
A solver for the solution of the free-space Poisson problem in three dimensions was presented and implemented in MATLAB for CPU and GPU usage. 
The method is spectral accurate and quasi linearly scaling. The computational domain can be a general rectangular box, where numerical 
experiments indicate acceptable error levels for moderately flat or prolongated domains and anisotropic densities.  The main computational tasks of the algorithm come from (zero-padded) FFTs. However, these operations are shown to be ideal for GPU acceleration, 
leading to a speedup factor compared to CPU of about $10$ for the Tesla GPU on the Vienna Scientific Cluster 3 (VSC3). The proposed approach and provided MATLAB implementation 
\footnote{Available on the author's webpage.}
is shown to be practically useful in terms of accuracy and efficiency. 
However, the code could also be understood as an easily readable prototype for translation to different programming languages.

\section*{Acknowledgments}
Financial support by the Austrian Science Fund (FWF) via the SFB ViCoM (grant F41) is acknowledged. 
The computations were achieved by using the Vienna Scientific Cluster 3 (VSC3).

\bibliographystyle{unsrtnat} 
\bibliography{refs}

\end{document}